\documentclass[12pt]{article}
\usepackage{graphicx}
\usepackage{amsmath}
\usepackage{gensymb}
\usepackage[backend=biber,maxbibnames=1,sorting=none]{biblatex}
\usepackage{booktabs}

\makeatletter 
\newcommand*{\rom}[1]{\expandafter\@slowromancap\romannumeral #1@}
\makeatother

\renewbibmacro{in:}{}
\DeclareFieldFormat{pages}{#1}%
\AtEveryBibitem{\clearfield{month}}
\AtEveryBibitem{\clearfield{day}}
\AtEveryCitekey{\clearfield{month}}
\AtEveryCitekey{\clearfield{day}}
\AtEveryBibitem{\clearfield{number}}
\AtEveryCitekey{\clearfield{number}}
\AtEveryBibitem{\clearfield{title}}
\AtEveryCitekey{\clearfield{title}}
\newcommand\pubnumber{}
\newcommand\pubdate{October 7, 2018}

\def\triumf{TRIUMF\\
4004 Wesbrook Mall, Vancouver, BC  V6T 2A3, Canada}

\def\Title#1{\begin{center} {\Large #1 } \end{center}}
\def\Author#1{\begin{center}{ \sc #1} \end{center}}
\def\Address#1{\begin{center}{ \it #1} \end{center}}

\newcommand\pubblock{\rightline{\begin{tabular}{l} \pubnumber\\
         \pubdate  \end{tabular}}}
\newenvironment{Abstract}{\begin{quotation}  }{\end{quotation}}
\newenvironment{Presented}{\begin{quotation} \begin{center} 
             PRESENTED AT\end{center}\bigskip 
      \begin{center}\begin{large}}{\end{large}\end{center} \end{quotation}}
\def\Acknowledgements{\bigskip  \bigskip \begin{center} \begin{large}
             \bf ACKNOWLEDGEMENTS \end{large}\end{center}}

\def\beq{\begin{equation}}
\def\eeq#1{\label{#1}\end{equation}}
\def\eeqn{\end{equation}}

\def\beqa{\begin{eqnarray}}
\def\eeqa#1{\label{#1}\end{eqnarray}}
\def\eeqan{\end{eqnarray}}

\let\bar=\overbar

\def\Dslash{\not{\hbox{\kern-4pt $D$}}}
\def\dslash{\not{\hbox{\kern-2pt $\del$}}}

\def\msb{{\bar{\ssstyle M \kern -1pt S}}}

\begin{document}
\begin{titlepage}
\pubblock

\vfill
\Title{$K^+ \to \pi^+ \nu \overline{\nu}$ -- NA62 First Result}
\vfill
\Author{Bob Velghe, for the NA62 Collaboration\footnotemark{}}
\footnotetext{\tiny{
R.~Aliberti, F.~Ambrosino, R.~Ammendola, B.~Angelucci, A.~Antonelli, G.~Anzivino, R.~Arcidiacono, M.~Barbanera, A.~Biagioni, L.~Bician
% ...
}}
\Address{\triumf}
\vfill
\begin{Abstract}
The CERN NA62 experiment uses a novel ``kaon decay-in-flight'' technique to observe $K^+ \to \pi^+ \nu \overline{\nu}$. The preliminary result based on the analysis of the full 2016 dataset will be presented. In agreement with the Standard Model prediction, one candidate was observed. Under the background hypothesis, an upper limit of $14 \times 10^{-10}$ at 95\,\% C.L. was placed on the branching ratio.
\end{Abstract}
\vfill
\begin{Presented}
CIPANP 2018\\
Palm Springs, CA, May 29 -- June 3, 2018
\end{Presented}
\vfill
\end{titlepage}
\def\thefootnote{\fnsymbol{footnote}}
\setcounter{footnote}{0}
Within the Standard Model (SM) framework, the $K^+ \to \pi^+ \nu \overline{\nu}$ decay is highly suppressed by the GIM mechanism \cite{Glashow1970} and the structure of the CKM matrix \cite{Kobayashi1973}. This fact, coupled to the clean theoretical prediction of the branching ratio, makes it an ideal probe for beyond the SM physics.

The branching ratio can be written in terms of the CKM parameters $V_{cb}$ and $\gamma$ \cite{Buras2015}:
\begin{equation}
	\mathcal{B}\left(K^+ \to \pi^+ \nu \overline{\nu}\right) = \left(8.39 \pm 0.30\right) \times 10^{-11} \left[\frac{\left|V_{cb}\right|}{40.7 \times 10^{-3}}\right]^{2.8} \left[\frac{\gamma}{73.2\degree} \right]^{0.74}~.
\label{eq:br_ckm}
\end{equation}
Using typical values for the parameters one finds \cite{pdg2017} 
\begin{equation}
\mathcal{B}\left(K^+ \to \pi^+ \nu \overline{\nu}\right) = \left(8.3 \pm 0.4\right) \times 10^{-11}~.
\nonumber
\end{equation}
Looking at Eq.~\ref{eq:br_ckm}, we see that the theoretical uncertainty amounts for 3.6\,\% of the error budget, the remaining is introduced by CKM parameters $V_{cb}$ and $\gamma$. 

Contrastingly, the current best experimental measurement, obtained the BNL E787 and E949 experiments using kaons decaying at rest, is \cite{Artamonov2009}
\begin{equation}
\mathcal{B}\left(K^+ \to \pi^+ \nu \bar{\nu}\right) = \left(17.3^{+11.5}_{-10.5}\right) \times 10^{-10}~,
\nonumber
\end{equation}
where the quoted errors are statistical.

Observable deviations from the SM branching ratio value have been predicted for a variety of scenarios, notably, $Z^\prime$ models \cite{Buras2015c}, Randall and Sandrum models \cite{Blanke2009}, Littlest Higgs models \cite{Blanke2016}, Supersymmetry \cite{Crivellin2011}, Lepton Flavour Violation \cite{Bordone2017}.

\section{CERN NA62 Experiment}
\begin{figure}
\includegraphics[width=\textwidth]{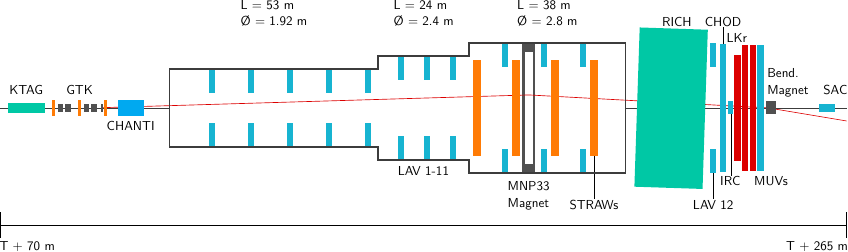}
\caption{
Diagram of the NA62 experiment. The primary SPS proton beam impinge a beryllium target installed 100\,m upstream of the experiment. A 75\,GeV/\textit{c} mixed hadron beam is then formed by a series of magnets and collimators. The beam particles are identified (KTAG) and their momentum are measured (GigaTracker) before entering the fiducial volume. Scintillators bars installed immediately after the last beam tracking station are used to detect unwanted beam interactions with the tracker sensors (CHANTI). Downstream, a spectrometer (STRAW) measures the momentum of the charged daughter particles. A RICH detector (RICH), a homogeneous electromagnetic calorimeter (LKr) and two sampling hadronic calorimeters (MUV 1 and 2) are used to select the pions. Two hodoscopes (CHOD) protect against possible beam interaction with the upstream detectors and complement the information on the event topology. Scintillator pads installed behind a iron wall are used to further reject muons (MUV3). Finally, a comprehensive system of photon detectors (SAC, IRC, LAV, LKr) provides hermetic photon coverage between 0 and 50\,mrad. The undecayed particles are travelling in a beam pipe and are not interacting with the downstream detectors. A multi-level, fully digital, trigger system selects events of interest based on energy deposition, and hit multiplicity and topologies. 
}
\label{fig:na62_layout}
\end{figure}
The NA62 collaboration aims to measure the $K^+ \to \pi^+ \nu \overline{\nu}$ branching ratio with an uncertainty of 10\,\%. 
The kaon decay-in-flight technique relies on the tracking of the kaon and its charged daughter particles, coupled with excellent timing capabilities needed to match the tracks in the high rate environment. Advanced particle identification and hermetic photon coverage allow to efficiently reject background while keeping the signal acceptance around 4\,\%. The experimental setup is briefly described in Fig.~\ref{fig:na62_layout}, further details can be found in \cite{Cortina2017}.

\section{Signal Selection}
\begin{figure}[h!]
    \begin{center}
			\includegraphics[width=0.7\textwidth]{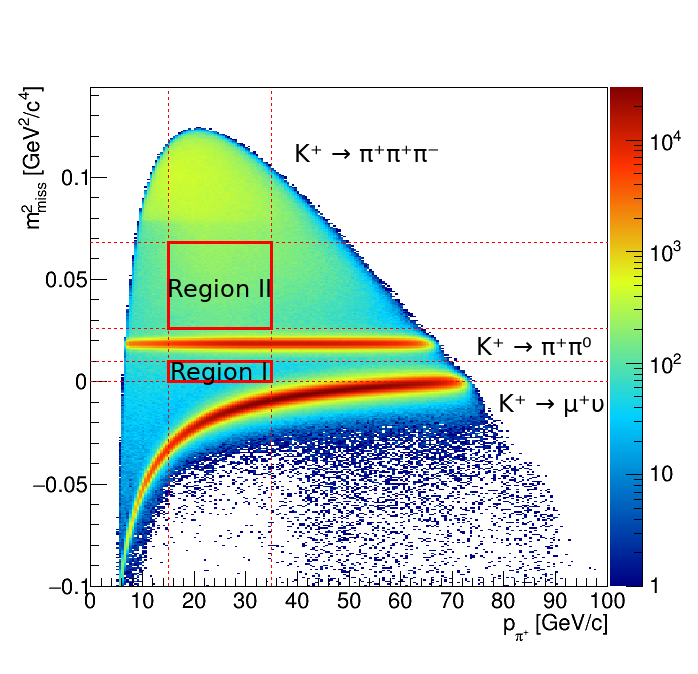}
    \end{center}
\caption{Plot of the missing mass squared, $m_\mathrm{miss}^2 \equiv \left(p_K - p_\pi\right)^2$, as a function of the STRAW momentum, assuming a pion track. The two signal regions are outlined by red boxes. }
\label{fig:kpinunu_kin}
\end{figure}
The presented results correspond to the study of the 2016 dataset \cite{Cortina2018b}.
Events with a single charged particle matched to a kaon track candidate are isolated. Cuts on the position of the reconstructed decay vertex restrict the acceptance to a pre-defined fiducial volume. To ensure a good separation of the pion and muon in the RICH and adequate photon rejection performances, the momentum of the downstream particle must lie between 15 and 35\,GeV/\textit{c}. The calorimeters are used to improve the pion identification. Events with in-time activity in the photon veto detectors or hints of beam related activity are discarded. 

As illustrated in Fig.~\ref{fig:kpinunu_kin}, two signal regions are defined in the missing mass squared STRAW momentum plane. The pion mass is assigned to the downstream particle when the missing mass squared is evaluated. The regions \rom{1} and \rom{2} extend, respectively, between the area populated by $K^+ \to \mu^+ \nu_\mu$ and $K^+ \to \pi^+ \pi^0$, and $K^+ \to \pi^+ \pi^0$ and $K^+ \to \pi^+ \pi^+ \pi^-$ decays.

\section{Preliminary Results}
\begin{table}
\caption{Key figures related to the 2016 dataset analysis.}
\label{tab:ses}
\begin{center}
\begin{tabular}{ll}
\toprule
$K^+ \to \pi^+ \nu \overline{\nu}$ acceptance & $\left(4.0 \pm 0.1\right) \times 10^{-2}$ \\ 
Accidental veto efficiency & $0.76 \pm 0.04$ \\
Trigger efficiency & $0.87 \pm 0.2$ \\
Kaon decays & $\left(1.21 \pm 0.02\right) \times 10^{11}$ \\
\bottomrule
\end{tabular}
\end{center}
\end{table}

\begin{table}[t]
\caption{Summary of the expected number of background events. The first and second quoted errors are, respectively, statistical and systematic. The number of expected $K^+ \to \pi^+ \nu \overline{\nu}$ candidates under the SM hypothesis is reported with the external uncertainty carried by the branching ratio prediction. The upstream backgrounds refer to beam related processes happening upstream of the fiducial volume.}
\label{tab:background_summary}
\begin{center}
	\begin{tabular}{ll}
	\toprule
	Process & Expected events \\
	\midrule
	$K^+ \to \pi^+ \pi^0 \left( \gamma \right)$ & $0.064 \pm 0.007 \pm 0.006$ \\
	Upstream backgrounds & $0.050^{+0.090}_{-0.030}$\\
	$K^+ \to \pi^+ \pi^- e^+ \nu$ & $0.018^{+0.024}_{-0.017} \pm 0.009$\\   
	$K^+ \to \pi^+ \pi^+ \pi^-$ & $0.002 \pm 0.001 \pm 0.002$  \\
	$K^+ \to \mu^+ \nu \left( \gamma \right)$ & $0.020 \pm 0.003 \pm 0.003$\\
	\midrule
	Total backgrounds & $0.15 \pm 0.09 \pm 0.01$ \\
	\midrule
	$K^+ \to \pi^+ \nu \overline{\nu}$ (SM) & $0.267 \pm 0.001 \pm 0.020 \pm 0.032_\mathrm{ext.}$ \\
	\bottomrule
	\end{tabular}
\end{center}
\end{table}

The analysis of the 2016 dataset yielded to a $K^+ \to \pi^+ \nu \overline{\nu}$ single event sensitivity (SES) of
\begin{equation}
    \left(3.15 \pm 0.01_\mathrm{stat.} \pm 0.24_\mathrm{syst.}\right) \times 10^{-10}~. \nonumber
\end{equation}
Key factors are detailed in Tab.~\ref{tab:ses}. The signal acceptance was derived from the Monte-Carlo (MC) simulation of the setup while the other factors were extracted from the data. The ``accidental veto efficiency'' refers to the signal acceptance reduction due to random activity in the vetoes detectors. The number kaon decays is inferred from a sample of $K^+ \to \pi^+ \pi^0$ selected with a control trigger.

A summary of the number of expected background events is given in Tab.~\ref{tab:background_summary}. Whenever possible, background estimates are data driven. The kinematic rejection factor is treated as independent of the muon and photon rejection. The contribution of the $K^+ \to \pi^+ \pi^0 \gamma$ radiative tail and $K^+ \to \pi^+ \pi^- e^+ \nu$ decay to the background were estimated with the MC. A bifurcation method \cite{Adler1997} was used to quantify the upstream backgrounds.

\begin{figure}
	\includegraphics[width=\textwidth]{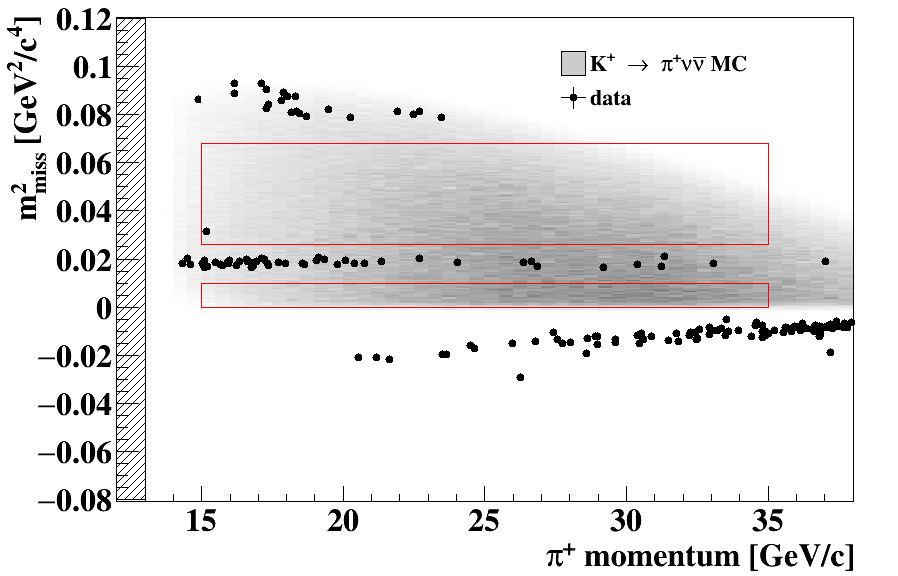}
	\caption{Overview of the \textit{cut-and-count} analysis of the 2016 dataset. One $K^+ \to \pi^+ \nu \overline{\nu}$ candidate was found in the region \rom{2}.}
\label{fig:pnn_unbox}
\end{figure}
To sum up, the first $K^+ \to \pi^+ \nu \overline{\nu}$ results obtained with a \textit{cut-and-count} approach are encouraging. Under the SM hypothesis, $0.267 \pm 0.001_\mathrm{stat.} \pm 0.02_\mathrm{syst.} \pm 0.032_\mathrm{ext.}$ events were expected in the combined signal regions. 
As shown in Fig.~\ref{fig:pnn_unbox}, after unblinding, one $K^+ \to \pi^+ \nu \overline{\nu}$ candidate was found in region \rom{2}.

If the candidate is assumed to be signal, the following branching ratio is found: 
\begin{equation}
\mathcal{B}\left(K^+ \to \pi^+ \nu \overline{\nu}\right) = 28^{+44}_{-23} \times 10^{-10}~68\,\%~\mathrm{C.L.}~. \nonumber
\end{equation}
On the contrary, if it is supposed to be background, one can derive an upper limit: 
\begin{equation}
\mathcal{B}\left(K^+ \to \pi^+ \nu \overline{\nu}\right) < 14 \times 10^{-10}~95\,\%~\mathrm{C.L.}~. \nonumber
\end{equation}

The 2017 and 2018 runs allowed us to multiply by twenty the amount of data collected compared to 2016. The reconstruction efficiency and overall acceptance will also be improved. Under the SM hypothesis, we expect of the order of twenty $K^+ \to \pi^+ \nu \overline{\nu}$ candidates before the accelerator maintenance stop planned in 2019 and 2020 (LHC Long Shutdown 2).

\Acknowledgements
This work was supported by the Natural Sciences and Engineering Research Council
of Canada and TRIUMF through a contribution from the National Research Council
of Canada.

\printbibliography

\end{document}